\documentclass{article} 
\usepackage{nips12submit_e,times}
\usepackage{epsfig,bm,epsf,latexsym,amsmath, amssymb, bbm, epsf, graphicx}

\newcommand{\argmin}[1]{\underset{#1}{\operatorname{argmin}}\;}
\newcommand{\sgn}{\operatorname{sgn}\;}
\newcommand{\la}{\left\langle}
\newcommand{\ra}{\right\rangle}

\newcommand{\be}{\begin{equation}}
\newcommand{\ee}{\end{equation}}
\newcommand{\bea}{\begin{eqnarray}}
\newcommand{\eea}{\end{eqnarray}}

\newcommand{\va}{{\boldsymbol{a}}}
\newcommand{\vb}{{\boldsymbol{b}}}

\newcommand{\vp}{{\boldsymbol{p}}}

\newcommand{\vh}{{\boldsymbol{h}}}

\newcommand{\vr}{{\boldsymbol{r}}}
\newcommand{\vs}{{\boldsymbol{s}}}

\newcommand{\vv}{{\boldsymbol{v}}}
\newcommand{\vw}{{\boldsymbol{w}}}
\newcommand{\vx}{{\boldsymbol{x}}}
\newcommand{\vy}{{\boldsymbol{y}}}

\newcommand{\vtheta}{{\boldsymbol{\theta}}}

\title{Compressive neural representation of sparse, high-dimensional probabilities}

\author{
xaq pitkow\\
Department of Brain and Cognitive Sciences\\
University of Rochester\\
Rochester, NY 14607 \\
\texttt{xpitkow@bcs.rochester.edu}
}

\nipsfinalcopy 

\begin{document}

\maketitle

\begin{abstract}
This paper shows how sparse, high-dimensional probability distributions could be represented by neurons with exponential compression. The representation is a novel application of compressive sensing to sparse probability distributions rather than to the usual sparse signals. The compressive measurements correspond to expected values of nonlinear functions of the probabilistically distributed variables. When these expected values are estimated by sampling, the quality of the compressed representation is limited only by the quality of sampling. Since the compression preserves the geometric structure of the space of sparse probability distributions, probabilistic computation can be performed in the compressed domain. Interestingly, functions satisfying the requirements of compressive sensing can be implemented as simple perceptrons. If we use perceptrons as a simple model of feedforward computation by neurons, these results show that the mean activity of a relatively small number of neurons can accurately represent a high-dimensional joint distribution implicitly, even without accounting for any noise correlations. This comprises a novel hypothesis for how neurons could encode probabilities in the brain.
\end{abstract}

\section{Introduction}

An arbitrary probability distribution over multiple variables has a parameter count that is exponential in the number of variables. Representing these probabilities can therefore be prohibitively costly. One common approach is to use graphical models to parameterize the distribution in terms of a smaller number of interactions. Here I consider an alternative approach. In many cases of interest, only a few unknown states have high probabilities while the rest have neglible ones; such a distribution is called `sparse'. I will show that sufficiently sparse distributions can be described by a number of parameters that is merely linear in the number of variables.

Until recently, it was generally thought that encoding of sparse signals required dense sampling at a rate greater than or equal to signal bandwidth. However, recent findings prove that it is possible to fully characterize a signal at a rate limited not by its bandwidth but by its information content \cite{Candes:2006p9443,Candes:2006p9438,Donoho:2006p9454,Candes:2011p9282} which can be much smaller. Here I apply such compression to sparse probability distributions over binary variables, which, after all, are just signals with some particular properties.

Traditional compressive sensing considers signals that lives in an $N$-dimensional space but have only $S$ nonzero coordinates in some basis. We say that such a signal is $S$-sparse. If we were told the location of the nonzero entries, then we would need only $S$ measurements to characterize their coefficients and thus the entire signal. But even if we don't know where those entries are, it still takes little more than $S$ linear measurements to perfectly reconstruct the signal. Furthermore, those measurements can be fixed in advance without any knowledge of the structure of the signal. Under certain conditions, these excellent properties can be guaranteed \cite{Candes:2006p9443,Candes:2006p9438,Donoho:2006p9454}.

The basic mathematical setup of compressive sensing is as follows. Assume that an $N$-dimensional signal $\vs$ has $S$ nonzero coefficients. We make $M$ linear measurements $\vy$ of this signal by applying the $M\times N$ matrix $A$:
\be
\vy=A\vs
\ee
We would then like to recover the original signal $\vs$ from these measurements. Under certain conditions on the measurement matrix $A$ described below, the original can be found perfectly by computing the vector with minimal $\ell_1$ norm that reproduces the measurements,
\be
\hat{\vs}=\argmin{\vs'}\|\vs'\|_{\ell_1}\text{ such that }A\vs'=\vy=A\vs
\label{eq:ClassicL1Recovery}
\ee
This is a powerful, practical result because (\ref{eq:ClassicL1Recovery}) can be solved efficiently \cite{Candes:2006p9443,Candes:2006p9438,Donoho:2006p9454,Donoho:2009p10095}.

Compressive sensing is generally robust to two deviations from this ideal setup. First, target signals may not be strictly $S$-sparse. However, they may be `compressible' in the sense that they are well approximated by an $S$-sparse signal. Signals whose rank-ordered coefficients fall off at least as fast as ${\rm rank}^{-1}$ satisfy this property \cite{Candes:2006p9438}. Second, measurements may be corrupted by noise with bounded amplitude $\epsilon$. Under these conditions, the error of the $\ell_1$-reconstructed signal $\hat{\vs}$ is bounded by the error of the best $S$-sparse approximation $\vs_S$ plus a term proportional to the measurement noise:
\be
\|\hat{\vs}-\vs\|_{\ell_2}\leq C_0 \|\vs_S-\vs\|_{\ell_2}/\sqrt{S}+C_1\epsilon
\label{eq:RobustReconstruction}
\ee
for some constants $C_0$ and $C_1$ \cite{Candes:2008p9630}.

Several conditions on $A$ have been used in compressive sensing to guarantee good performance \cite{Candes:2006p9438,Candes:2011p9282,Kueng:2012p9339,Calderbank:2010p9985,Gurevich:2009p10035}. Modulo various nuances, they all essentially ensure that most or all relevant sparse signals lie sufficiently far from the null space of $A$: It would be impossible to recover signals in the null space since their measurements are all zero and cannot therefore be distinguished. The most commonly used condition is the Restricted Isometry Property (RIP), which says that $A$ preserves $\ell_2$ norms of all $S$-sparse vectors within a factor of $1\pm\delta_S$ that depends on the sparsity,
\be
(1-\delta_S)\|\vs\|_{\ell_2}\ \leq\ \|A\vs\|_{\ell_2}\ \leq\ (1+\delta_S)\|\vs\|_{\ell_2}
\ee
If $A$ satsifies the RIP with small enough $\delta_S$, then $\ell_1$ recovery is guaranteed to succeed. For random matrices whose elements are independent and identically distributed Gaussian or Bernoulli variates, the RIP holds as long as the number of measurements $M$ satisfies
\be
M\geq C S\log{N/S}
\ee
for some constant $C$ that depends on $\delta_S$ \cite{Candes:2008p9630}. No other recovery method, however intractable, can perform substantially better than this \cite{Candes:2008p9630}.

\section{Compressing sparse probability distributions}

Compressive sensing allows us to use far fewer resources to accurately represent high-dimensional objects if they are sufficiently sparse. Even if we don't ultimately intend to reconstruct the signal, the reconstruction theorem described above (\ref{eq:RobustReconstruction}) ensures that we have implicitly represented all the relevant information. This compression proves to be extremely useful when representing multivariate joint probability distributions, whose size is exponentially large even for the simplest binary states.

Consider the signal to be a probability distribution over an $n$-dimensional binary vector $\vx\in\{-1,+1\}^n$, which I will write sometimes as a function $p(\vx)$ and sometimes as a vector $\vp$ indexed by the binary state $\vx$. I assume $\vp$ is sparse in the canonical basis of delta-functions on each state, $\delta_{\vx,\vx'}$. The dimensionality of this signal is $N=2^n$, which for even modest $n$ can be so large it cannot be represented explicitly.

The measurement matrix $A$ for probability vectors has size $M \times 2^n$. Each row corresponds to a different measurement, indexed by $i$. Each column corresponds to a different binary state $\vx$. This column index $\vx$ ranges over all possible binary vectors of length $n$, in some conventional sequence. For example, if $n=3$ then the column index would take the 8 values
$$
\vx\in\left\{-\!-\!-\ ;\ -\!-\!+\ ;\ -\!+\!-\ ;\ -\!+\!+\ ;\ +\!-\!-\ ;\ +\!-\!+\ ;\ +\!+\!-\ ;\ +\!+\!+\right\}
$$
Each element of the measurement matrix, $A_i(\vx)$, can be viewed as a function applied to the binary state. When this matrix operates on a probability distribution $p(\vx)$, the result $\vy$ is a vector of $M$ expectation values of those functions, with elements
\be
y_i = A_i\vp=\sum_\vx A_{i}(\vx)p(\vx) =\left\langle A_i(\vx)\right\rangle_{p(\vx)}
\ee
For example, if $A_i(\vx)=x_i$ then $y_i=\la x_i\ra_{p(\vx)}$ measures the mean of $x_i$ drawn from $p(\vx)$.

For suitable measurement matrices $A$, we are guaranteed accurate reconstruction of $S$-sparse probability distributions as long as the number of measurements is
\be
M\geq O(S\log{N/S})=O(S n-S\log{S})
\ee
Note that the exponential size of the probability vector, $N=2^n$, is cancelled by the logarithm. For distributions with a fixed sparseness $S$, the required number of measurements per variable, $M/n$, is then independent of the number of variables.\footnote{Depending on the problem, the number of significant nonzero entries $S$ may grow with the number of variables. This growth may be fast (e.g. the number of possible patterns grows as $e^n$) or slow (e.g. the number of possible translations of a given pattern grows only as $n$).}

In many cases of interest it is impractical to calculate these expectation values directly: Recall that the probabilities may be too expensive to represent explicitly in the first place. One remedy is to draw $T$ samples $\vx_t$ from the distribution $p(\vx)$, and use a sum over these samples to approximate the expectation values,
\be
y_i\approx \frac{1}{T}\sum_t A_i(\vx_t)\hspace{.35in}\vx_t \sim p(\vx)
\label{eq:CompressiveSampling}
\ee

The probability $\hat{p}(\vx)$ estimated from $T$ samples has errors with variance $p(\vx)(1-p(\vx))/T$, which is bounded by $1/4T$. This allows us to use the performance limits from robust compressive sensing, which according to (\ref{eq:RobustReconstruction}) creates an error in the reconstructed probabilities that is bounded by
\be
\|\hat{\vp}-\vp\|_{\ell_2}\leq C_0\|\vp_S-\vp\|_{\ell_2}+\frac{C_1}{\sqrt{T}}
\ee
where $\vp_S$ is a vector with the top $S$ probabilities preserved and the rest set to zero.

\subsection{Measurements by random perceptrons}

In compressive sensing it is common to use a matrix with independent Bernoulli-distributed random values, $A_i(\vx)\sim\mathcal{B}(\tfrac{1}{2})$, which guarantees $A$ satisfies the RIP \cite{Mendelson:2006p10101}. Each row of this matrix represents all possible outputs of an arbitrarily complicated Boolean function of the $n$ binary variables $\vx$.

\begin{figure}[hbtp]
\centering
\includegraphics*[width=5.5in]{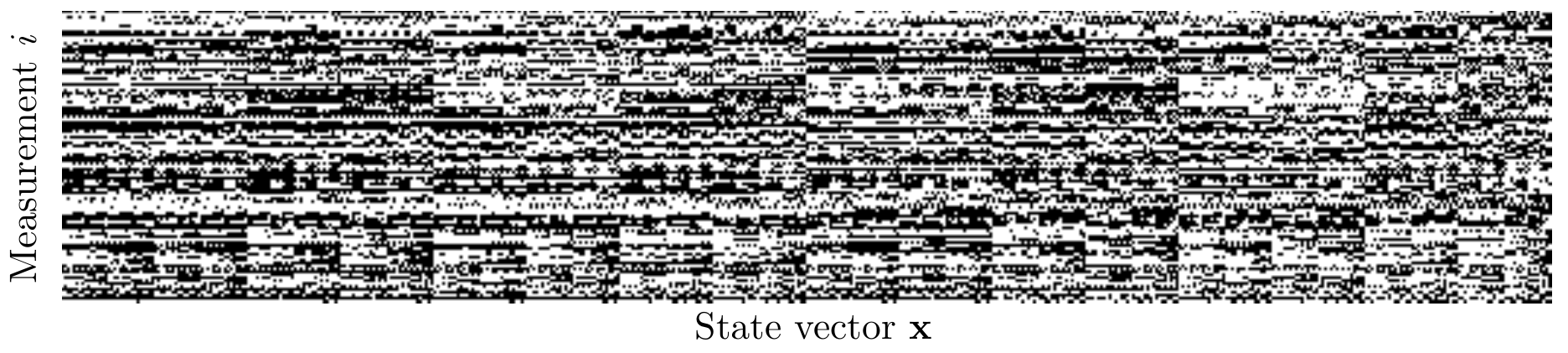}
\caption{Example measurement matrix $A_i(\vx)$ for $M=100$ random perceptrons applied to all $2^9$ possible binary vectors of length $n=9$.}
\label{fig:RandomPerceptron}
\end{figure}

Biological neural networks would have great difficulty computing such arbitrary functions in a simple manner. However, neurons can easily compute a large class of simpler boolean functions, the perceptrons. These are simple threshold functions of a weighted average of the input
\be
A_i(\vx)=\sgn{\!\left( \sum\nolimits_j\!W_{ij}x_j-\theta_j\right)}
\label{eq:RandomPerceptron}
\ee
where $W$ is an $M\times n$ matrix. Here I take $W$ to have elements drawn randomly from a standard normal distribution, $W_{ij}\sim\mathcal{N}(0,1)$, and call the resultant functions `random perceptrons'. An example measurement matrix for random perceptrons is shown in Figure \ref{fig:RandomPerceptron}. These functions are readily implemented by individual neurons, where $x_j$ is the instantaneous activity of neuron $j$, $W_{ij}$ is the synaptic weight between neurons $i$ and $j$, and the ${\rm sgn}$ function approximates a spiking threshold at $\theta$.

The step nonlinearity $\sgn$ is not essential, but {\em some} type of nonlinearity is. Using a simple linear function of the states, $A=W\vx$, would result in measurements $\vy=A\vp=W\la\vx\ra$. This provides at most $n$ linearly independent measurements of $p(\vx)$, even when $M>n$. In most cases this is not enough to adequately capture the full distribution.

Although the dimensionality of $W$ is merely $M\times n$, which is much smaller than the $2^n$-dimensional space of probabilities, (\ref{eq:RandomPerceptron}) can generate $O(2^{n^2})$ distinct perceptrons \cite{Irmatov:2009p8882}. By including an appropriate threshold, a perceptron can assign any individual state $\vx$ a positive response and assign a negative response to every other state. This shows that random perceptrons generate the canonical basis and can thus span the space of possible $p(\vx)$. In what follows, I assume that $\vtheta=0$ for simplicity.

Below I present empirical evidence that a small number of random perceptrons preserves the information about sparse distributions. In the Appendix I show that random perceptrons with zero threshold are incoherent with the canonical basis, and the rows of the random perceptron measurement matrix are asymptotically orthogonal in the limit of large $n$. Random perceptrons thus satisfy the requirements for RIPless compressive sensing \cite{Candes:2011p9282}. Present research is directed toward deriving the condition number of these matrices for finite $n$, in order to provide rigorous bounds on the number of measurements required in practice.

\section{Experiments}

\subsection{Fidelity of compressed sparse distributions}
\label{sec:reconstruction}

To test random perceptrons in compressive sensing of probabilities, I generated sparse distributions using small Boltzmann machines \cite{ackley1985learning}, and compressed them using random perceptrons driven by samples from the Boltzmann machine. Performance was then judged by $\ell_1$ reconstructions.

In a Boltzmann Machine, binary states $\vx$ occur with probabilities given by the Boltzmann distribution
\be
p(\vx)\propto e^{-E(\vx)}
\label{eq:BoltzmannDistribution}
\ee
for an energy function
\be
E(\vx)=-\vb^\top\! \vx-\vx^\top\! J\vx
\ee
determined by biases $\vb$ and pairwise couplings $J$. Sampling from this distribution can be accomplished by running Glauber dynamics \cite{Glauber:1963p294}, at each time step turning a unit on with probability $p(x_i=+1|\vx_{\setminus i})=1/(1+e^{-\Delta E})$, where $\Delta E=E(x_i=+1,\vx_{\setminus i})-E(x_i=-1,\vx_{\setminus i})$. Here $\vx_{\setminus i}$ is the vector of all components of $\vx$ except the $i$th.

For simulations I distinguished between two types of units, hidden and visible, $\vx=(\vh,\vv)$. On each trial I first generated a sample of all units according to (\ref{eq:BoltzmannDistribution}). I then fixed only the visible units and allowed the hidden units to fluctuate according to the conditional probability $p(\vh|\vv)$ to be represented. This probability is given again by the Boltzmann distribution, now with energy function
\be
E(\vh|\vv)=-(\vb_h-J_{hv}\vv)^\top\vh-\vh^\top\! J_{hh}\vh
\ee
All bias terms $\vb$ were set to zero, and all pairwise couplings $J$ were random draws from a zero-mean normal distribution, $J_{ij}\sim\mathcal{N}(0,\frac{1}{3})$. Experiments used $n$ hidden and $n$ visible units, with $n\in\{8,10,12\}$. This distribution of couplings produced sparse posterior distributions whose rank-ordered probabilities fell faster than ${\rm rank}^{-1}$ and were thus compressible \cite{Candes:2006p9438}.

\begin{figure}[hbtp]
\centering
\includegraphics*[width=5.5in]{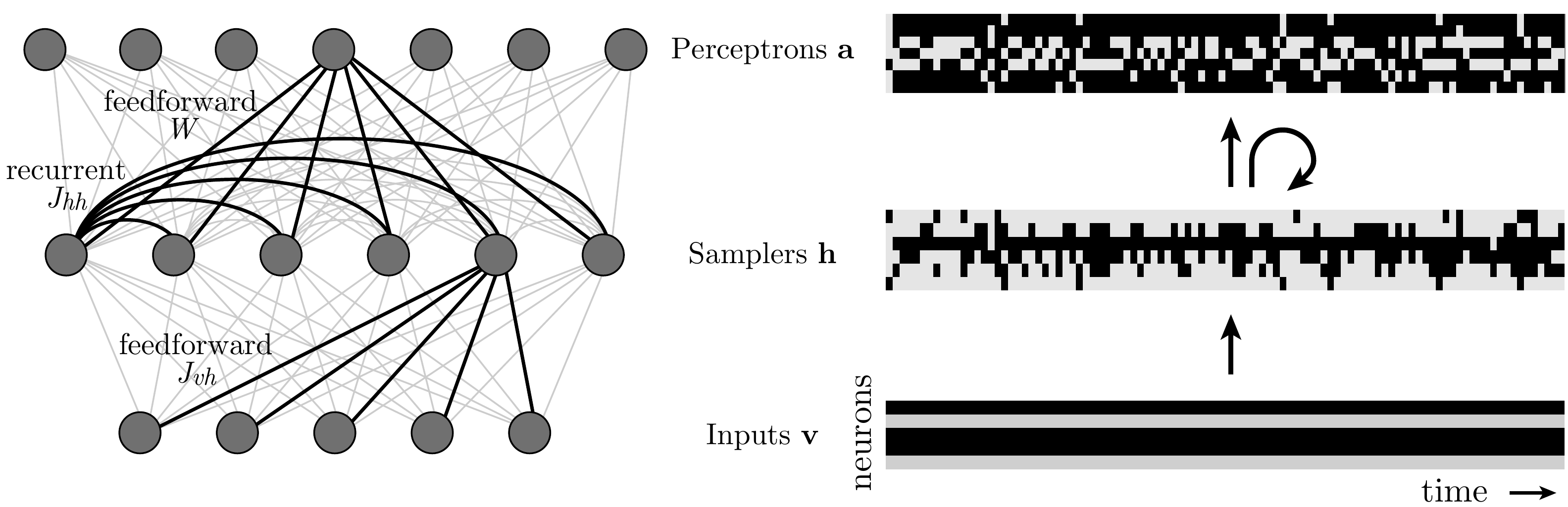}
\caption{Compressive sensing of a probability distribution by model neurons. Left: a neural architecture for generating and then encoding a sparse, high-dimensional probability distribution. Right: activity of each population of neurons as a function of time. Sparse posterior probability distribution are generated by a Boltzmann Machine with visible units $\vv$ (Inputs), hidden units $\vh$ (Samplers), feedforward couplings $J_{vh}$ from visible to hidden units, and recurrent connections between hidden units $J_{hh}$. The visible units' activities are fixed by an input. The hidden units are stochastic, and sample from a probability distribution $p(\vh|\vv)$. The samples are recoded by feedforward weights $W$ to random perceptrons $\va$. The mean activity $\vy$ of the time-dependent perceptron responses captures the sparse joint distribution of the hidden units.}
\label{fig:NeuralNetwork}
\end{figure}

The compression was accomplished by passing the hidden unit activities $\vh$ through random perceptrons $\va$ with weights $W$, according to $\va=\sgn{\!(W\vh)}$. These perceptron activities fluctuate along with their inputs. The mean activity of these perceptron units compressively senses the probability distribution according to (\ref{eq:CompressiveSampling}). This process of sampling and then compressing a Boltzmann distribution can be implemented by the simple neural network shown in Figure \ref{fig:NeuralNetwork}.

\begin{figure}[hbtp]
\centering
\includegraphics*[width=5.5in]{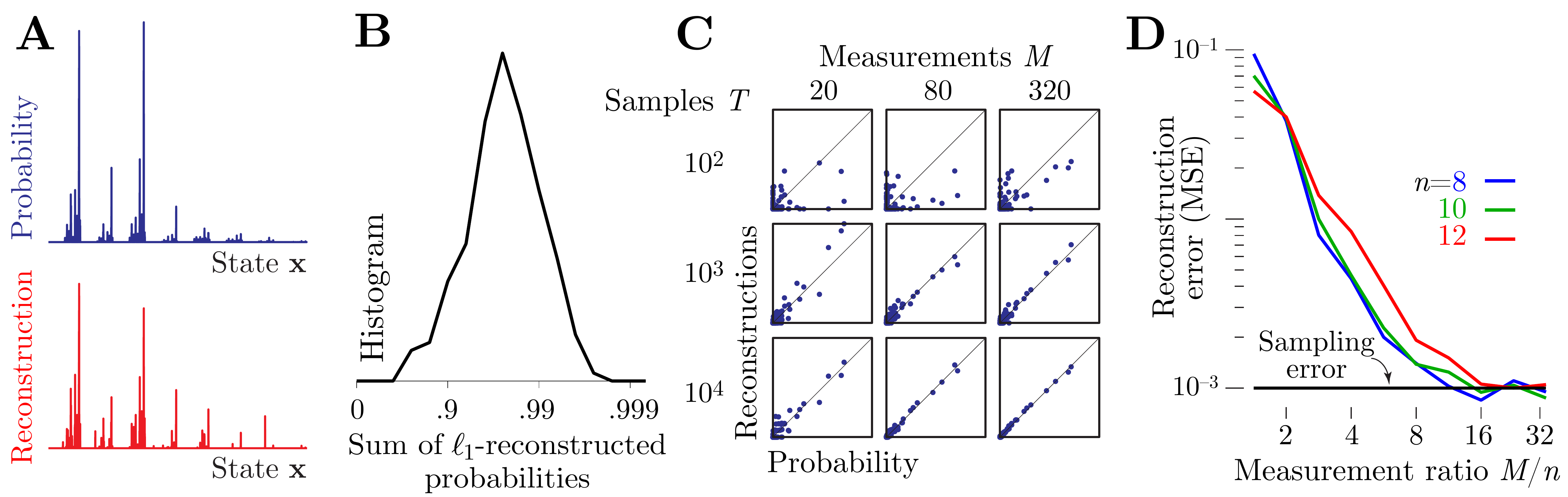}
\caption{Reconstruction of sparse posteriors from random perceptron measurements. ({\bf A}) A sparse posterior distribution over 10 nodes in a Boltzmann machine is sampled 1000 times, fed to 50 random perceptrons, and reconstructed by nonnegative $\ell_1$ minimization. ({\bf B}) A histogram of the sum of reconstructed probabilities reveals the small shortfall from a proper normalization of 1. ({\bf C}) Scatter plots show reconstructions versus true probabilities. Each box uses different numbers of compressive measurements $M$ and numbers of samples $T$. ({\bf D}) With increasing numbers of compressive measurements, the mean squared reconstruction error falls to $1/T=10^{-3}$, the limit imposed by finite sampling.}
\label{fig:Reconstructions}
\end{figure}

We are not ultimately interested in reconstruction of the large sparse distribution, but rather its compressed representation. Nonetheless, reconstruction is useful to show that the information has been preserved. I reconstruct sparse probabilities using nonnegative $\ell_1$ minimization with measurement constraints \cite{Yang:2011p250,Zhang:2010}, minimizing
\be
\|\vp\|_{\ell_1}+\lambda \|A\vp-\vy\|_{\ell_2}^2
\ee
where $\lambda$ is a regularization parameter that was set to $2T$ in all simulations. Reconstructions were quite good, as shown in Figure \ref{fig:Reconstructions}. Even with far fewer measurements than signal dimensions, reconstruction accuracy is limited only by the sampling of the posterior. Enough random perceptrons do not lose any available information.

In the context of probability distributions, $\ell_1$ reconstruction has a serious flaw: All distributions have the same $\ell_1$ norm: $\|\vp\|_{\ell_1}=\sum_\vx p(\vx)=1$! To minimize the $\ell_1$ norm, therefore, the estimate will not be a probability distribution. Nonetheless, the individual probabilities of the most significant states are accurately reconstructed, and only the highly improbable states are set to zero. Figure \ref{fig:Reconstructions}B shows that the shortfall is small: $\ell_1$ reconstruction recovers over 90\% of the total probability mass.

\subsection{Preserving computationally important relationships}

\begin{figure}[hbtp]
\centering
\includegraphics*[width=5.5in]{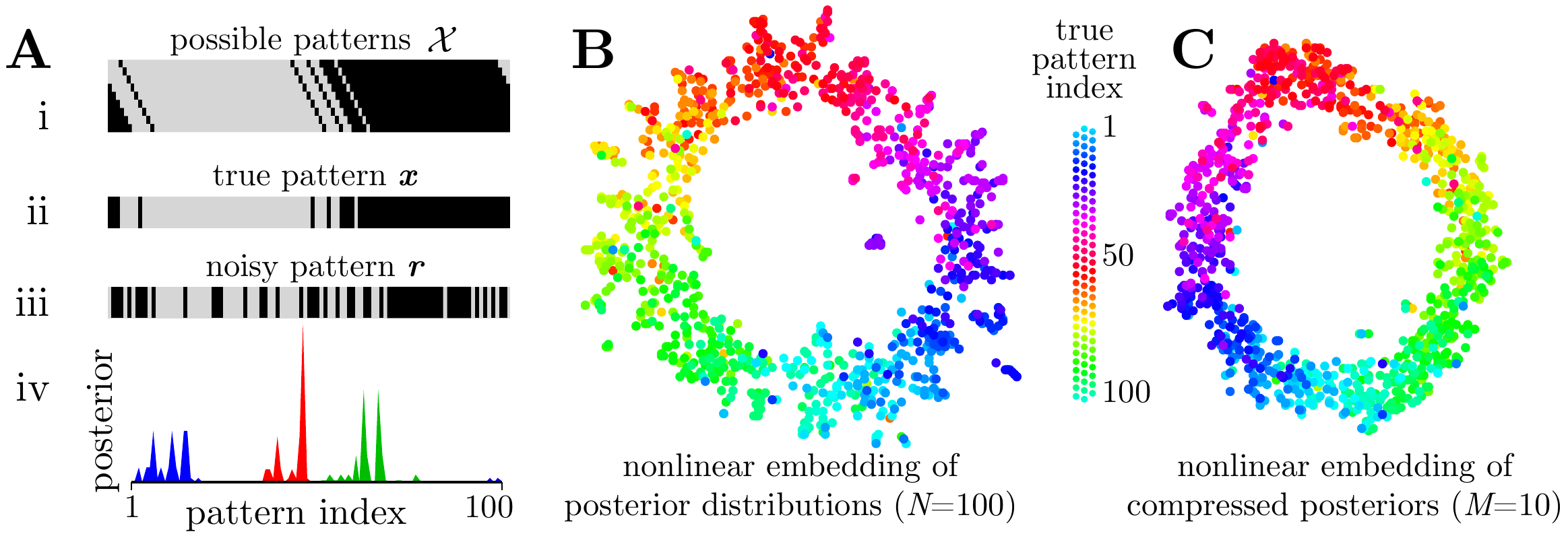}
\caption{Nonlinear embeddings of a family of probability distributions with a translation symmetry. ({\bf A}) The process of generating posterior distributions: (i) A set of 100 possible patterns is generated as cyclic translations of a binary pattern (only 9 shown). With uniform probability, one of these patterns is selected (ii), and a noisy version is obtained by randomly flipping bits with probability 0.35 (iii). From such noisy patterns, an observer can infer posterior probability distributions over possible inputs (iv). ({\bf B}) The set of posteriors from 1000 iterations of this process is nonlinearly mapped \cite{Maaten:2008p9376} from 100 dimensions to 2 dimensions. Each point represents one posterior and is colored according to the actual pattern from which the noisy observations were made. The permutation symmetry of this process is revealed as a circle in this mapping. ({\bf C}) This circular structure is retained even after each posterior is compressed into the mean output of 10 random perceptrons.}
\label{fig:NonlinearEmbedding}
\end{figure}

There is value in being able to compactly represent these high-dimensional objects. However, it would be especially useful to perform probabilistic computations using these representations, such as marginalization and evidence integration. Since marginalization is a linear operation on the probability distribution, this is readily implementable in the linearly compressed domain. In contrast, evidence integration is a multiplicative process acting in the canonical basis, so this operation will be more complicated after the linear distortions of compressive measurement $A$. Nonetheless, such computations should be feasible as long as the informative relationships are preserved in the compressed space: Similar distributions should have similar compressive representations, and dissimilar distributions should have dissimilar compressive representations. In fact, that is precisely the guarantee of compressive sensing: topological properties of the underlying space are preserved in the compressive domain \cite{Baraniuk:2010p8689}. Figure \ref{fig:NonlinearEmbedding} illustrates how not only are individual sparse distributions recoverable despite significant compression, but the topology of the set of all such distributions is retained.

For this experiment, an input $\vx$ is drawn from a dictionary of input patterns $\mathcal{X}\subset\{+1,-1\}^n$. Each pattern in $\mathcal{X}$ is a translation of a single binary template $\vx^0$ whose elements are generated by thresholding a noisy sinusoid (Figure \ref{fig:NonlinearEmbedding}A): $x^0_j=\sgn[4\sin{(2\pi j/n)}+\eta_j]$ with $\eta_j\sim\mathcal{N}(0,1)$. On each trial, one of these possible patterns is drawn randomly with equal probability $1/|\mathcal{X}|$, and then is measured by a noisy process that randomly flips bits with a probability $\eta=0.35$ to give a noisy pattern $\vr$. This process induces a posterior distribution over the possible input patterns
\begin{align}
p(\vx|\vr)&=\frac{1}{Z}p(\vx)p(\vr|\vx)=\frac{1}{Z}p(\vx)\prod_ip(r_i|x_i)\\
&=\frac{1}{Z}p(\vx)\eta^{N-h(\vx,\vr)}(1-\eta)^{h(\vx,\vr)}
\end{align}
where $h(\vx,\vr)$ is the Hamming distance between $\vx$ and $\vr$. This posterior is nonzero for all patterns in the dictionary. The noise level and the similarities between the dictionary elements together control the sparseness.

1000 trials of this process generates samples from the set of all possible posterior distributions. Just as the underlying set of inputs has a translation symmetry, the set of all possible posterior distributions has a cyclic permutation symmetry. This symmetry can be revealed by a nonlinear embedding \cite{Maaten:2008p9376} of the set of posteriors into two dimensions (Figure \ref{fig:NonlinearEmbedding}B).

Compressive sensing of these posteriors by 10 random perceptrons produces a much lower-dimensional embedding that preserves this symmetry. Figure \ref{fig:NonlinearEmbedding}C shows that the same nonlinear embedding algorithm applied to the reduced representation, and one sees the same topological pattern. In compressive sensing, similarity is measured by Euclidean distance. When applied to probability distributions it will be interesting to examine instead how well information-geometric measures like the Kullback-Leibler divergence are preserved under this dimensionality reduction \cite{CarterRFH11}.

\section{Discussion}

Probabilistic inference appears to be essential for both animals and machines to perform well on complex tasks with natural levels of ambiguity, but it remains unclear how the brain represents and manipulates probability. Present population models of neural inference either struggle with high-dimensional distributions \cite{Ma:2006p839} or encode them by hard-to-measure high-order correlations \cite{Berkes:2011p9843}. Here I have proposed an alternative mechanism by which the brain could efficiently represent probabilities: random perceptrons. In this model, information about probabilities is compressed and distributed in neural population activity. Amazingly, the brain need not measure any correlations between the perceptron outputs to capture the joint statistics of the sparse input distribution. Only the mean activities are required. Figure \ref{fig:NeuralNetwork} illustrates one network that implements this new representation, and many variations on this circuit are possible.

Successful encoding in this compressed representation requires that the input distribution be sparse. Posterior distributions over sensory stimuli like natural images are indeed expected to be highly sparse: the features are sparse \cite{Olshausen:1996p9890}, the prior over images is sparse \cite{Stephens:2008p9760}, and the likelihood produced by sensory evidence is usually restrictive, so the posteriors should be even sparser. Still, it will be important to quantify just how sparse the relevant posteriors are under different conditions. This would permit us to predict how neural representations in a fixed population should degrade as sensory evidence becomes weaker.

Brains appear to have a mix of structure and randomness. The results presented here show that purely random connections are sufficient to ensure that a sparse probability distribution is properly encoded. Surprisingly, more structured connections cannot allow a network with the same computational elements to encode distributions with substantially fewer neurons, since compressive sensing is already nearly optimal \cite{Candes:2008p9630}. On the other hand, some representational structure may make it easier to perform computations later. Note that unknown randomness is not an impediment to further processing, as reconstruction can be performed even without explicit knowledge of random perceptron measurement matrix \cite{Isely:2010p9708}.

Even in the most convenient representations, inference is generally intractable and requires approximation. Since compressive sensing preserves the essential geometric relationships of the signal space, learning and inference based on these relationships may be no harder after the compression, and could even be more efficient due to the reduced dimensionality. Biologically plausible mechanisms for implementing probabilistic computations in the compressed representation is important work for the future.

\appendix
\section*{Appendix: Asymptotic orthogonality of random perceptron matrix}

To evaluate the quality of the compressive sensing matrix, we need to ensure that $S$-sparse vectors are not projected to zero by the action of $A$. Here we show that the inner products between the columns of $\hat{A}=A/\sqrt{M}$ are concentrated around zero by computing their mean and variance over the ensemble of $W_{ij}\sim\mathcal{N}(0,1)$ and for random pairs of binary states $\vx$ and $\vx'$. For compactness I will write $\vw_i$ for the $i$th row of the perceptron weight matrix $W$.

First I compute the mean and variance of the mean inner product $\la{C}_{\vx\vx'}\ra_W$ between columns of $\hat{A}$ for a given pair of states $\vx\neq\vx'$:
\be
\la{C}_{\vx\vx'}\ra_W=\la\sum\nolimits_i\hat{A}_i(\vx)\hat{A}_i(\vx')\ra_W=\frac{1}{M}\sum\nolimits_i \la\sgn{\!\!\left(\vw_i\!\cdot\!\vx\right)}\sgn{\!\!\left(\vw_{i}\!\cdot\!\vx'\right)}\ra_{\!W}
\label{eq:MeanInnerProduct}
\ee
and since the different $\vw_{i}$ are independent, this implies that
\be
\la{C}_{\vx\vx'}\ra_W=\la\sgn{\!\!\left(\vw_i\!\cdot\!\vx\right)}\sgn{\!\!\left(\vw_{i}\!\cdot\!\vx'\right)}\ra_{\!W}
\label{eq:MeanInnerProduct2}
\ee
The $n$-dimensional half-space in $W$ where $\sgn{\!\!(\vw_i\cdot\vx)}=+1$ intersects with the corresponding half-space for $\vx'$ in a wedge-shaped region with an angle of $\theta=\cos^{-1}(\vx\cdot\vx'/\|\vx\|_{\ell_2}\|\vx'\|_{\ell_2})$. This angle is related to the Hamming distance $h=h(\vx,\vx')$:
\be
\theta(h)=\cos^{-1}(\vx\cdot\vx'/n)=\cos^{-1}(1-2h/n)
\ee
The signs of $\vw_i\!\cdot\!\vx$ and $\vw_i\!\cdot\!\vx'$ agree within this wedge region and its reflection about $W=0$, and disagree in the supplementary wedges. The mean inner product is therefore
\begin{align}
\la C_{\vx\vx'}\ra_W=&
(+1)P\left[\,\sgn{\!\!(\vw_i\!\cdot\!\vx)}=\sgn{\!\!(\vw_i\!\cdot\!\vx')}\right]+\\
&(-1)P\left[\,\sgn{\!\!(\vw_i\!\cdot\!\vx)}\neq\sgn{\!\!(\vw_i\!\cdot\!\vx')}\right]\\
=&1-2\theta(h)/\pi
\label{eq:MeanInnerProduct3}
\end{align}
The variance of $C_{\vx\vx'}$ caused by variability in $W$ is given by
\begin{align}
V_{\vx\vx'}&=\la C_{\vx\vx'}^2\ra_W-\la C_{\vx\vx'}\ra^2_W\\
&=\sum_{i=j}\la\hat{A}_i^2(\vx)\hat{A}_i^2(\vx')\ra_W+\sum_{i\neq j}\la\hat{A}_i(\vx)\hat{A}_i(\vx')\hat{A}_j(\vx)\hat{A}_j(\vx')\ra_W-\la C_{\vx\vx'}\ra^2_W\\
&=\sum_i \la\frac{\sgn{\!\!\left(\vw_i\!\cdot\!\vx\right)^2}}{M}\frac{\sgn{\!\!\left(\vw_{i}\!\cdot\!\vx'\right)^2}}{M}\ra_{\!W}\!\!\!\!+\sum_{i\neq j}\la\frac{\sgn{\!\!\left(\vw_{i}\!\cdot\!\vx\right)}}{\sqrt{M}}\frac{\sgn{\!\!\left(\vw_{i}\!\cdot\!\vx'\right)}}{\sqrt{M}}\ra_W^2\!\!\!-\la C_{\vx\vx'}\ra^2_W\\
&=\frac{1}{M}+\frac{M^2-M}{M^2}(1-2\theta(h)/\pi)^2-\la C_{\vx\vx'}\ra_W^2\\
&=\frac{1}{M}\left(1-\left(1-\tfrac{2}{\pi}\theta(h(\vx,\vx'))\right)^2\right)
\label{eq:MeanInnerProduct}
\end{align}
This variance falls with $M$, so for large numbers of measurements $M$ the inner products between columns concentrates around the various state-dependent mean values (\ref{eq:MeanInnerProduct3}).

Next I consider the diversity of inner products for different pairs $(\vx,\vx')$ of binary state vectors, in the limit of large $M$ so that the diversity is dominated by variations over the different pairs. The mean inner product depends only on the Hamming distance $h$ between $\vx$ and $\vx'$, which for sparse signals with random support has a binomial distribution, $p(h)=\binom{n}{h}2^{-n}$ with mean $n/2$ and variance $n/4$. Designating by an overbar the average over randomly chosen states, the mean $\overline{C}$ and variance $\overline{\delta C^2}$ of the inner product is
\begin{align}
\overline{C}&=\overline{\la C_{\vx\vx'}\ra_W}=1-\tfrac{2}{\pi}\overline{\cos^{-1}(1-\tfrac{2h}{n})}=0\\
\overline{\delta C^2}&=\overline{\delta h^2}\left(\frac{\partial C}{\partial h}\right)^2=\frac{n}{4}\frac{16}{\pi^2 n^2}=\frac{4}{\pi^2 n}
\end{align}
This proves that in the limit of large $n$ and $M$, the columns of the random perceptron measurement matrix have inner products that concentrate around 0. The Grammian of $A$ is just a matrix of many such inner products, and is orthonormal almost surely, $A^\top\! A\to I$. Consequently, with enough measurements a sparse signal can be recovered perfectly according to RIPless compressive sensing \cite{Candes:2011p9282}. Future work will determine how this Grammian matrix behaves for finite $n$ and $M$, which will determine the number of measurements required in practice to capture a signal of a given sparseness.

\subsection*{Acknowledgments}

Thanks to Alex Pouget, Jeff Beck, Shannon Starr, and Carmelita Navasca for helpful conversations.

\footnotesize

\bibliographystyle{plos}
\bibliography{SparseDistributions_Bib2}

\end{document}